\pdfoutput=1
\documentclass[11pt]{article}
\usepackage{graphicx}
\usepackage{adjustbox}
\usepackage{booktabs}   
\usepackage{subcaption} 
\usepackage{algpseudocode}
\usepackage{algorithm}
\usepackage{bm}
\usepackage[colorinlistoftodos]{todonotes}
\usepackage{times}
\usepackage{latexsym}
\usepackage{tabulary}
\usepackage[T1]{fontenc}
\usepackage[utf8]{inputenc}
\usepackage{hyperref}
\usepackage{microtype}
\pdfoutput=1
\usepackage[]{EACL2023}

\usepackage{inconsolata}

\title{GAP-Gen: Guided Automatic Python Code Generation}

\renewcommand{\thefootnote}{\fnsymbol{footnote}}

\author{Junchen Zhao \footnote[1]{Equally contributed\label{eq}} \hspace{30pt} Yurun Song \footnotemark
 \hspace{30pt}  Junlin Wang \hspace{30pt}  Ian G. Harris \\ 
\vspace{5pt}
junchez3@uci.edu\hspace{20pt} yuruns@uci.edu\hspace{20pt}  junliw1@uci.edu\hspace{20pt} harris@ics.uci.edu \\
\vspace{5pt}
        University of California, Irvine }

\begin{document}
\maketitle 
\footnotetext{* Equally contributed}
\renewcommand{\thefootnote}{\arabic{footnote}}
\setcounter{footnote}{0}
\begin{abstract}
Automatic code generation from natural language descriptions can be highly beneficial during the process of software development. In this work, we propose GAP-Gen, a \textbf{G}uided \textbf{A}utomatic \textbf{P}ython \textbf{C}ode \textbf{G}eneration method based on Python syntactic constraints and semantic constraints. We first introduce Python syntactic constraints in the form of \textbf{Syntax-Flow}, which is a simplified version of Abstract Syntax Tree (AST) reducing the size and high complexity of real python AST but maintaining crucial syntactic information of Python code. In addition to Syntax-Flow, we introduce \textbf{Variable-Flow}  which abstracts variable and function names consistently throughout the code. In our work, rather than pre-training, we focus on modifying the fine-tuning process which reduces computational requirements but retains high generation performance on automatic Python code generation task. GAP-Gen fine-tunes the transformer-based language models T5 and CodeT5 using the Code-to-Docstring datasets CodeSearchNet, CodeSearchNet AdvTest and Code-Docstring-Corpus from EdinburghNLP.  Our experiments show that GAP-Gen achieves better results on automatic Python code generation task than previous works. Our implementation is available on the github\footnote{\url{https://github.com/Rain9876/Auto-Code-Generator}}.


\end{abstract}

\begin{figure*}[ht] 
\centering
\begin{subfigure}{\textwidth}
\vspace{-50pt}
  {
    \includegraphics[width=\textwidth, height=8cm]{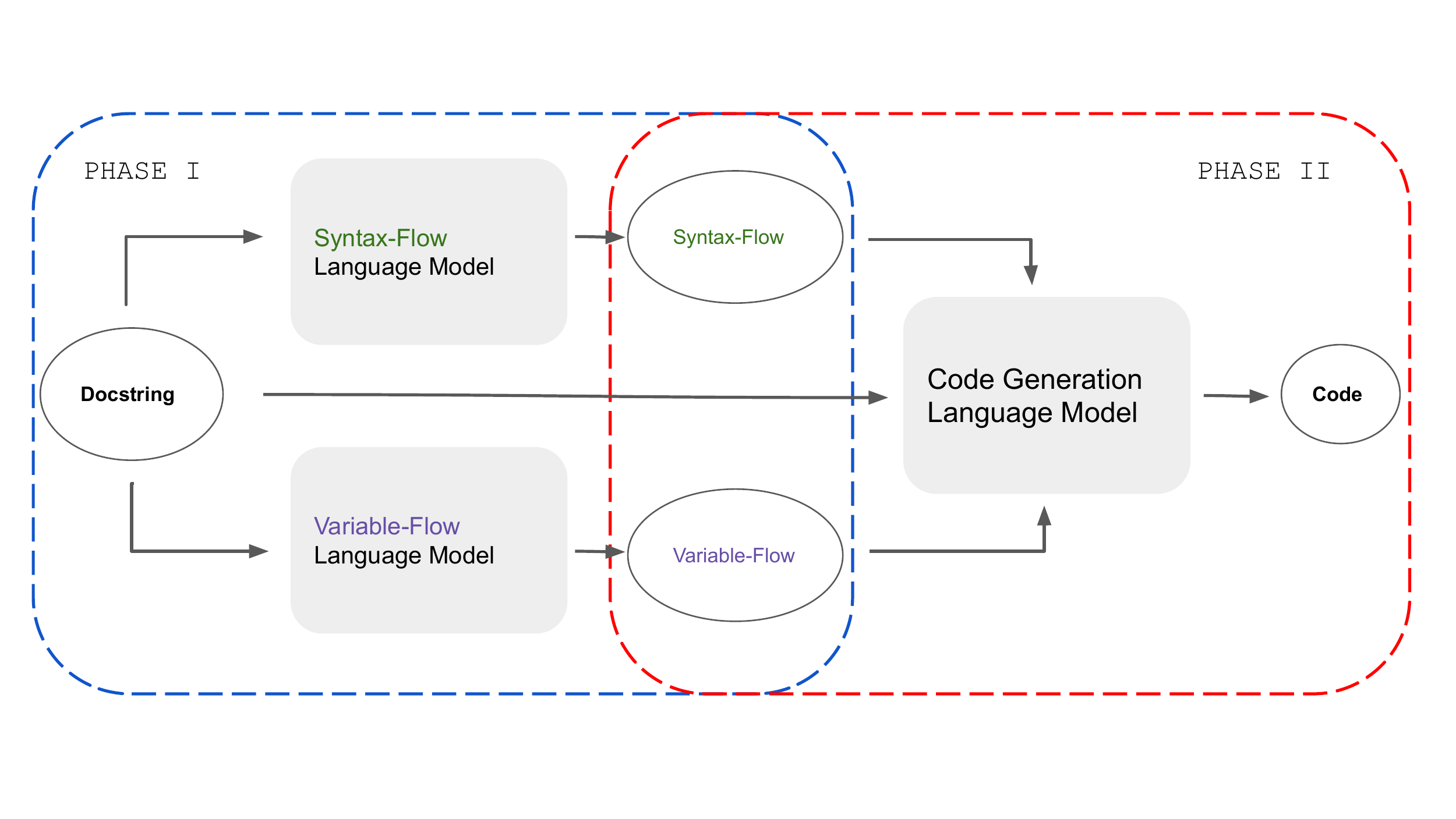}
}
\label{fig:overview:orig}
  \end{subfigure}
     \vspace{-50pt}
  \caption{\textbf{An overview of the proposed approach} Phase I has two language models generating Syntax-Flow and Variable-Flow. In Phase II, another language model encodes these two types of information as well as the docstring to generate code.}
  \label{fig:overview}
\end{figure*}
\section{Introduction}

With billions of people relying on software for their everyday work and life, developers face an ongoing challenge to create programs efficiently.
One potential solution to this challenge is to use human descriptions to generate the corresponding source code. By using this approach, developers can write software specifications in natural language, which are then translated into code through code generation mechanism.
Early attempts to tackle this problem were rule-based, identifying syntactic patterns in text and using handcrafted rules to map the patterns to code. Methods used to recognize
syntactic structure include regular patterns (\citealp[]{conf/sigmod/GulwaniM14}; \citealp[]{conf/aaai/KateWM05}; \citealp[]{Le2013SmartSynthSS}) and parse
trees produced using context-free grammars (\citealp[]{conf/aaai/KateWM05}; \citealp[]{Le2013SmartSynthSS}; \citealp[]{Ballard1979ProgrammingIN}; \citealp[]{Price2000NaturalJavaAN}). Several
previous approaches convert a sentence into a formal statement
by mapping verbs to functions in the formal language, and
mapping the objects of the verb in the sentence to function arguments in the formal language (\citealp[]{Ballard1979ProgrammingIN}; \citealp[]{Price2000NaturalJavaAN}; \citealp[]{Little2006TranslatingKC}). For example, the
sentence, “Add r1 to r2” might be mapped to add(r1, r2) in
a procedural language. The problem of finding the objects of
the verb to use as function arguments is simple if the sentence
structure is strictly limited. Several approaches use regular
expressions \citep{Le2013SmartSynthSS} or context-free grammars \citep{conf/aaai/KateWM05} to identify the
objects in the sentence.


More recent approaches are data-driven and
leverage machine learning methods, e.g., \citep{Desai2016ProgramSU} uses a Naive
Bayesian Classifier to map English words to a domain-specific
language, and \citep{Quirk2015LanguageTC} learns production rules for a semantic
parser. \citep{Rahit2019MachineTF} uses a Long-Short Term Memory
(LSTM) Recurrent Neural Network (RNN) architecture to
implement their neural machine translation approach . Work
presented in \citep{Ling2016LatentPN} introduces the Latent Predictor Network
(LPN) architecture which treats code generation as a sequence-to-sequence modeling problem. \citep{Yin2017ASN} builds upon
this approach by leveraging the grammar model of the target
language as prior knowledge.

Research presented in (\citealp[]{Clement2020PyMT5MT}; \citealp[]{Feng2020CodeBERTAP}; \citealp[]{Lu2021CodeXGLUEAM}) introduced transformer-based language model pre-training methods to map the natural language semantic with the code. Although these works have relatively good performance on source code generation task, they require high computational resources, which are difficult to acquire. They also usually consider the code as a sequence of tokens (\citealp[]{Feng2020CodeBERTAP}; \citealp[]{kanade2020learning}; \citealp[]{Lu2021CodeXGLUEAM}) and ignore either the source code's syntactic-level or semantic-level information, which could improve the language models' code understanding capability, during their pre-training process. 


In this work, we present GAP-Gen, a method to improve automatic Python source code generation from natural language description. Our GAP-Gen is fine-tuning of the pre-trained T5-English \citep{raffel2020exploring} and CodeT5 \citep{wang2021codet5} language models that employ Syntax-Flow and Variable-Flow as guidance and has shown on being able to understand the relationship between natural language description and Python code from syntactic and semantic level of the Python code. 

Our GAP-Gen training pipeline is composed of two phases. As shown in Figure \ref{fig:overview}, \textbf{Phase I}, fine-tunes our pre-trained language model for the purpose of generating Python code's syntactic constraints and semantic-level structure, the Syntax-Flow and the Variable-Flow. \textbf{Phase II}, fine-tunes a separate language model by encoding natural language description of the code, the generated code syntactic constrains (Syntax-Flow) and abstracted variable names (Variable-Flow) from Phase I to generate Python code. By doing so, language models fine-tuned with GAP-Gen training pipeline are able to surpass many previous works' performances, which rely on the pre-training process of language models without considering code's syntactic and semantic information.

Our \textbf{main contributions} are:
\begin{itemize}
\item We introduce Syntax-Flow and demonstrate the importance of the source code's syntactic information in the automatic Python code generation task.
\item We show that abstracting variable and function names through Variable-Flow is effective in maintaining the naming semantics of the code.
\item We achieve high performance on automatic Python source code generation task without language model pre-training.

\end{itemize}


\section{Related Works}

\textbf{Language Models for Programming Languages.} Transformer-based language models that utilize attention mechanisms have been dominating NLP benchmarks (\citealp[]{Vaswani2017AttentionIA}; \citealp[]{Wang2018GLUEAM}). The novel attention-based message passing techniques plus multi-task pre-training \citep{Devlin2019BERTPO} have been through extensive studies. This leads to a deeper understanding of the representational power of transformer-based models (\citealp[]{Ethayarajh2019HowCA}; \citealp[]{Kovaleva2019RevealingTD}; \citealp[]{Jain2019AttentionIN}).

At the same time, transformer-based auto-regressive language models consisting of encoder/decoder demonstrate stellar performances on many NLP generative tasks (\citealp[]{Radford2019LanguageMA}; \citealp[]{Lewis2020BARTDS}; \citealp[]{Raffel2020ExploringTL}). These tasks include but not limited to story generation \citep{See2019DoMP}, dialogue \citep{Budzianowski2019HelloIG}, summarization \citep{Lewis2020BARTDS}, Entity Retrieval \citep{DeCao2021AutoregressiveER}, Question Answering \citep{Guu2020REALMRL}, and so on. Similar advances have also been made in Programming Language relevant tasks. 

Programming language generation tasks, although not considered as natural language generation tasks, have been demonstrated to have great results when they are modeled similarly as natural language generation tasks. \citep{Feng2020CodeBERTAP} pre-trains on Mask Language Modeling (MLM) and replaced-token detection for code understanding tas. In \citep{Liu2020MultitaskLB}, the authors develop a code completion transformer-based model by jointly predicting the probability and type of the next token. For the task of code summarization, transformer-based models outperform the other neural approaches (\citealp[]{Yu2020TowardsCC}; \citealp[]{Ahmad2020ATA}); \citealp[]{10.1145/3368089.3417058}; \citealp[]{Liu2020MultitaskLB}) use GPT and UniLM respectively for code completion.
More related to our work, (\citealp[]{Husain2019CodeSearchNetCE}; \citealp[]{Clement2020PyMT5MT}) explore pre-training methodologies for learning better structural and syntactical information for automatic code generation. Moreover, (\citealp[]{wang2021codet5}; \citealp[]{guo2021graphcodebert}) incorporates Variable-Flows and identifier information into their pre-training process for better code generation performance.



\textbf{Guided Text Generative Models.} Generative modeling is powerful but often falls short in many conditions. The behavior of auto-regressive language models cannot be explicitly controlled, and was shown to be very easy to degenerate (\citealp[]{Holtzman2020TheCC}; \citealp[]{Welleck2020NeuralTG}; \citealp[]{Meister2020IfBS}). This is also the case for code generation. This prompts researchers to combat this issue by looking at either the training time or the decoding time. Work in \citep{Fan2018HierarchicalNS} constrains the sample space to top-k tokens in the softmax logistics to avoid introducing highly unlikely tokens. \citep{Holtzman2020TheCC} instead restricts the sampling space to the smallest set of space above some probability mass. Using simple decoding variants is lightweight to implement, but does not change the predicted likelihood of each token. \citep{Welleck2020NeuralTG} argues that the likelihood objectives is at fault, and proposes unlikelihood training objective, which forces lower probabilities on unlikely generations. 

Furthermore, practitioners have also injected priors or structural information into the language model for better generation. (\citealp[]{Zhang2020OptimizingTF}; \citealp[]{Lagutin2021ImplicitUT}) utilizes policy learning to control model behaviors. However, this approach suffers from high variance \citep{Choshen2020OnTW}. Recently work in story generation (\citealp[]{Yao2019PlanAndWriteTB}; \citealp[]{Rashkin2020PlotMachinesOG}; \citealp[]{GoldfarbTarrant2020ContentPF}) uses a plotline/storyline as an intermediate state for generation. This alleviates the language modeling tasks and sets up the model to better learn the structure of the stories. Being motivated by story generation, our work injects syntactic and semantic structural information in a setup that is similar to this line of works. For the code generation task, we utilize our proposed Syntax-Flow and Variable-Flow  as the intermediate state to help language model better understand code's syntactic and semantic structure information and improve its performance.

\begin{figure*}[ht] 
\centering
\includegraphics[width=\textwidth]{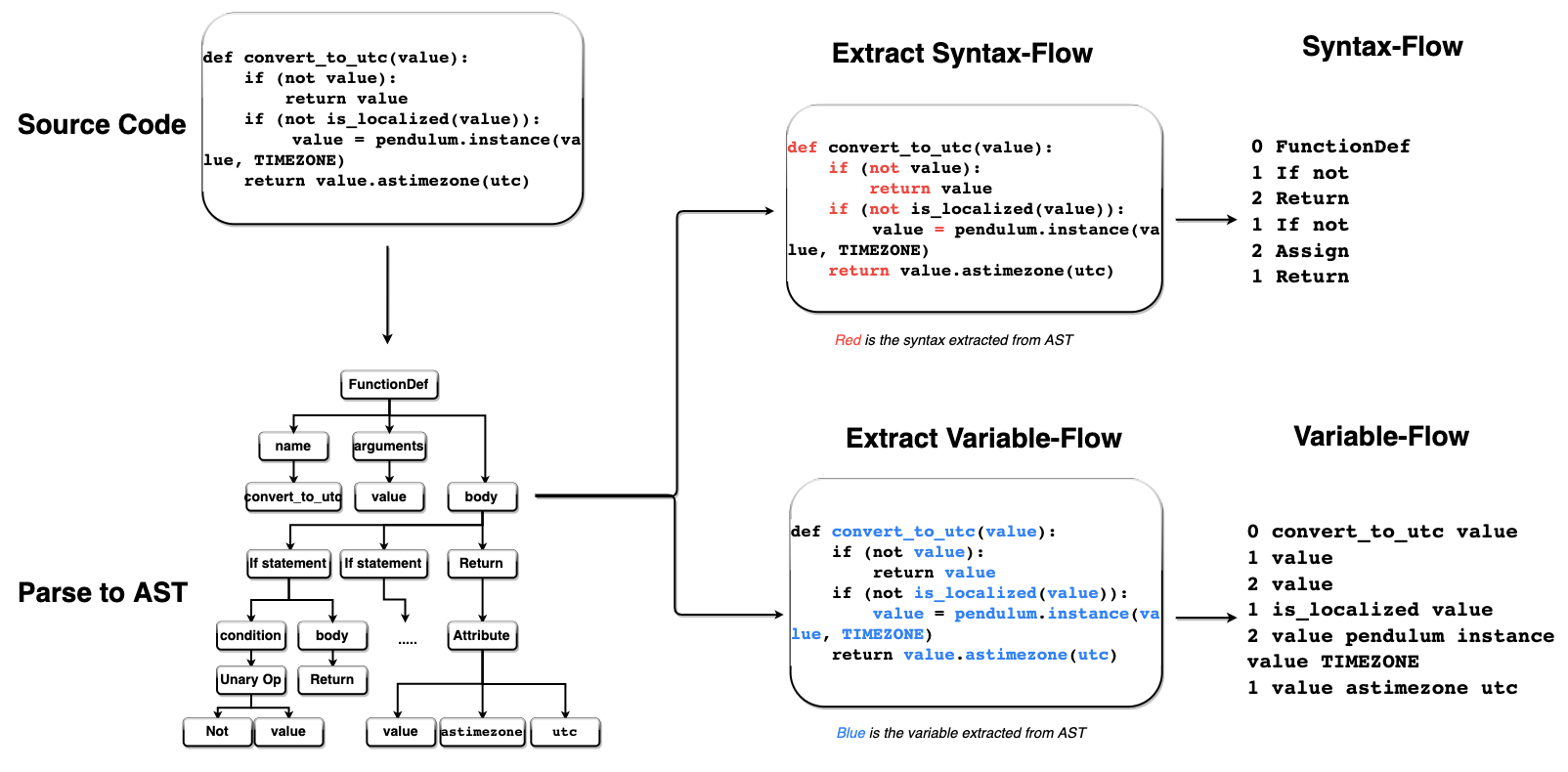}
\vspace{-10pt}
\caption{\textbf{An overview of the Syntax-Flow and Variable-Flow Generation }. The numbers refer to the number of indentations (4 spaces) required at the beginning of the command line.  For Syntax-Flow, statements (STMTs) immediately follow the required indentation and are then followed by several built-in expressions (Exprs). In contrast, Variable Flow follows the function and variable names.}
\label{fig:phase_I}
\end{figure*}

\section{Method}
In this section, we describe our method by introducing Syntax-Flow and Variable-Flow. Then we present the generation process of Syntax-Flow and Variable-Flow. Finally, we present the Python code generation process guided by Syntax-Flow and Variable-Flow. 
\subsection{Syntax-Flow} 
Unlike other methods that generate code directly from source input with pre-training, our approach works in a pipeline by generating the structure of the code as an intermediate state first and then generating the detailed code using the code structure. 



Procedural structure can be expressed formally as an Abstract Syntax Tree (AST). An AST contains two major components: \textbf{STMT (Statement)} and \textbf{Expr (Expression)}. STMT describes the general structure of code including the high-level Python code syntactic constraints. Expr is the detailed content of the code, mainly including the function variables and operations. Additionally, there are some special components in an AST such as the exceptional handler, import alias, arguments, etc.




Due to the AST's formality and rich expressiveness regarding the syntactic information of code, there are works that generate an AST first and then use it to aid code generation, such as (\citealp[]{Yin2017ASN}; \citealp[]{Ling2016LatentPN}), but these works usually require composite model architecture changes. Also, the AST is too complex for models to directly generate information. The length of ASTs typically makes it impractical to directly input them into transformer-based language models due to their input sequence length limitation. As a result, we propose a simplified version of an AST, namely Syntax-Flow.

Instead of using the entire tree structure of the AST, we only extract crucial information including Indentation, STMT, and some parts of Expr. By doing so, we reduce the complexity of AST but retain its crucial syntactic structure of Python code, and is small enough to be compatible with transformer-based language models.


In our proposed Syntax-Flow, there are three critical components: \textbf{Indentation}, \textbf{STMT} and \textbf{Default Functions}. These three components are viewed as invariants which means that these components are kept unchanged for maintaining code's correct functionality.


\subsection{Variable-Flow} 

Variable-Flow is another indispensable component in automatic code generation task. It can be effectively applied to maintain the naming semantics of the code during the code generation process. \cite{wang2021codet5,guo2021graphcodebert} use Variable-Flow during their pre-training process and achieve good performances on programming language-relevant tasks. In their works, they extract function variables names as Variable-Flow which is integrated into their pre-training process for improving language models' capability on understanding the code semantic structure. 


In our work, rather than extract variable names only as Variable-Flow, our Variable-Flow contains \textbf{Indentation}, \textbf{variable names}, and \textbf{function names}. Variable names and function names are uniformly free to change. In other words, Python code's functionality remains correct regardless of the changes in these two Variable-Flow components. Therefore, compared with Syntax-Flow, our Variable-Flow contains variant components and is more dynamic. 




 

\subsection{Phase I - Generation of Syntax-Flow and Variable-Flow }


\subsubsection{Generation of Syntax-Flow}
Figure~\ref{fig:phase_I} shows the Syntax-Flow and Variable-Flow generation pipeline. With regard to Syntax-Flow, the numbers represent the number of indentations (4 spaces) required at the beginning of the command line. Then, a statement is immediately next to the Indentation, followed by several built-in expressions.
This feature is extracted from Python code through AST syntax visitor method, a method to go through every detail of the AST nodes recursively and extracts all necessary nodes for use, such as FunctionDef, STMT, exception handler etc, with the count of indentation at the same time. 


The simplified generation process of Syntax-Flow is shown in Algorithm 1 Appendix~\ref{sec:appendix}. For each line of source code, we generate one line of Syntax-Flow as you can see in Figure~\ref{fig:phase_I}. Formally we denote the source code to be $y = (y_1, y_2,...,y_n)$, and let $E = [e_1,e_2,...,e_L]$ be the list of indexes of the newline character. Hence, $y_{e_1}$ is the first line break, and $Y_1 = (y_1,...,y_{e_1})$ is the first line of the source code, $Y_2 = (y_{e_1+1},...,y_{e_2})$ the second and so on. Then for each such line of the code, we generate a pair $a = (t, c)$. $t$ is the indentation of the current line of code or number of tabs, and $c$ is the code logic which includes control flow or function definitions. In other words, we are looking for, 
\begin{equation}
p(t_i,c_i|Y_i) = p(a_i|Y_i) 
\end{equation}

where $i$ denotes the $i$th line. Both properties are derived from an AST, which is generated by a standard toolkit. For more detailed steps, refer to Algorithm 1 in Appendix~\ref{sec:appendix}.

\subsubsection{Syntax-Flow Language Model}
To better learn and utilize the syntactical information of the source code, we use the Syntax-Flow language model to first encode docstrings and then generate Syntax-Flow. Here we use a pre-trained auto-regressive language model. We do not do any additional pre-training, so computing resource is restricted to a manageable amount. As shown in Algorithm 2 in Appendix~\ref{sec:appendix}, to fine-tune the language model, we first generate AST from the ground golden source code $|\bm{y}=(\bm{y_1},\bm{y_2},\dots,\bm{y_n})$. Then we transform the AST of the source code to Syntax-Flow in a deterministic process,

\begin{equation}
a = \textproc{SynParse}( \textproc{AST-Parse}(y))
\end{equation}

where \textproc{SynParse} stands for Syntax-Flow Parse. Both \textproc{SynParse} and \textproc{ASTParse} are deterministic functions that generate the Syntax-Flow $a$. We take this as our true reference and model the process as a standard generative task $P_{LM_S}(\hat{a}_{i} | x,\hat{a}_{1} ... \hat{a}_{i-1})$, namely, 
\begin{equation}
\hat{a} = LM_S(x)
\end{equation}

where $x$ is the input (docstring for code generation). During inference, given a docstring, this language model is able to generate Syntax-Flow directly for the latter use. 

\subsubsection{Generation of Variable-Flow } 
We define the format of Variable-Flow in our work similar to that of Syntax-Flow as shown in Figure~\ref{fig:phase_I}. For each line of code, it has an indentation $t$ followed by $V = [v_1,...,v_j]$. $V$ is the list of Variable-Flow which can be either variable names or function names. Multiple variable names can exist in the same line. Its sequential nature alleviates language models like T5 during the generation process. Similar to the setup of Syntax-Flow, we are looking for 
\begin{equation}
p(t_i,V_i|Y_i) = p(b_i|Y_i) 
\end{equation}

where $Y_i$ is the $i$th line of source code and $b_i = (t_i,V_i)$. 
\subsubsection{Variable-Flow Language Model}
We generate Variable-Flow from source code $\bm{y}=(\bm{y_1},\bm{y_2},\dots,\bm{y_n})$. Then we can safely extract Variable-Flow determinedly from AST:
\begin{equation}
b = \textproc{VarParse}(\textproc{ASTParse}(y))
\end{equation}
where \textproc{VarParse} stands for Variable-Flow Parse. We take this as our true reference and fine-tune the Variable-Flow Language Model on the source code and 
Variable-Flow pairs. Specifically, we are modeling $P_{LM_V}(\hat{b}_{i} | x,\hat{b}_{1} ... \hat{b}_{i-1})
$. Hence,
\begin{equation}
\hat{b} = LM_V(x)
\end{equation}

for the $i$th line. At inference time, the model is expected to generate Variable-Flow for the latter models to encode.


\subsection{Phase II -- Generation of Code}
Code generation is built on the same language model as Syntax-Flow and Variable-Flow language model. However, unlike the generation of Syntax-Flow or Variable-Flow, an issue in the code generation task is that the length of code description is usually much shorter than the length of generated code.
For example, in the CodeSearchNet dataset, many function code data length is over 128 tokens while the description only has an average length is about 50 tokens per sequence. This means that the input information is limited and not enough to generate plausible code unless the language model is available to have more prepared features during the code generation process.
For this reason, we use the information generated from Phase I as intermediate features to guide the language model generating Python code in Phase II.

\subsubsection{Guided Code Generation Language Model} 

Our Code Generation Language Model depends on the docstring and the corresponding Syntax-Flow and the Variable-Flow. The language model is obtained from a pre-trained auto-regressive language model T5. In our work, we use the T5-based language model as our guided Code Generation Language Model $LM_G $.
\begin{equation}
\hat{y} = LM_G(x, LM_S(x), LM_V(x))
\end{equation}
The Guided Code Generation Language Model takes in the input docstrings $x$ as well as the outputs of the Syntax-Flow Language Model and Variable-Flow Language Model. 

\begin{table*}[ht]
\centering
\begin{tabular}{|l|c|c|c|c|}
\hline
                 & Rouge1-F1                    & Rouge2-F1                   & RougeL-F1   & BLEU  \\ \hline
CSN Syntax-Flow   & 49.1                      & 35.8                          & 47.7    & 12.7 \\ 
CSN Variable-Flow & 36.7                        & 15.7                        & 33.7    & 11.4 \\ \hline
CDC Syntax-Flow   & 51.8                       & 41.4                         & 50.4    & 15.2 \\ 
CDC Variable-Flow & 37.4                        & 18.9                        & 34.8    & 11.9 \\ \hline
Adv Syntax-Flow   & 50.4                        & 36.9                        & 48.9    & 13.6 \\ 
Adv Variable-Flow & 37.3                        & 15.6                        & 34.0    & 11.1 \\ \hline  
\end{tabular}
\caption{The results of Syntax-Flow and Variable-Flow generation for all three datasets in Phase I with T5. The performance is evaluated through Rouge and BLUE.}
\label{table::phase1}
\end{table*}

\begin{table*}[ht]
\centering
\begin{tabular}{|l|c|c|c|c|c|}
\hline
             & Rouge1-F1 & Rouge2-F1 & RougeL-F1  & BLEU    & CodeBLEU    \\ \hline
CSN          & 31.1  & 12.1  & 27.9   & 21.2   & 22.1       \\ \hline
CDC          & \textbf{32.3}  & \textbf{15.7}  & \textbf{29.3}   & \textbf{22.6}   & \textbf{22.4}       \\ \hline
Adv          & 29.8  & 11.0  & 26.7   & 20.7   & 20.9       \\ \hline
\end{tabular}
\caption{The results of Python code generation for CSN, CDC and Adv in Phase II using GAP-Gen pipeline with T5. The performance is evaluated through Rouge, BLUE and CodeBLEU.}
\label{Phase_II}
\end{table*}
\begin{table*}[ht]
\centering

\begin{tabular}{|l|c|c|c|c|c|}
\hline
                                   & Rouge1-F1    & Rouge2-F1     & RougeL-F1  & BLEU   & CodeBLEU \\ \hline
GPT2 \citep{Clement2020PyMT5MT}    & 20.9      & 7.6        & 21.9    & 2.8    & --       \\ \hline
PyMT5 \citep{Clement2020PyMT5MT}   & 28.4      & 13.5       & 24.8    & 8.6   & --       \\ \hline
T5           & 30.4   & 11.7   & 27.4   & 20.7  & 21.7         \\ \hline
GAP-Gen T5  & \textbf{31.1}  & \textbf{12.1}  & \textbf{27.9}   & \textbf{21.2}   & \textbf{22.1}         \\ \hline
CodeT5 \citep{wang2021codet5}      & 34.6     & 14.6      & 30.2        & 21.6      &  23.4      \\ \hline
GAP-Gen CodeT5   & \textbf{35.1}     & \textbf{14.9}      & \textbf{30.6}        & \textbf{22.3}      & \textbf{24.1}        \\ \hline
\end{tabular}
\caption{The results of GAP-Gen with other models fine-tuning on CSN datasets for Python code generation task. We report the Rouge, BLUE and CodeBLEU score for all different models, where GAP-Gen T5 and GAP-Gen CodeT5 are the models built on the T5 and CodeT5 model separately using GAP-Gen pipeline.
}
\label{other_results}
\end{table*}

\section{Experiment}
In this section, we present our experiment in detail. First, we introduce the datasets we use and our data processing approach in our experiment. Then, we present our experimental setup. Finally, we introduce our evaluation metrics in the last subsections.
 
\subsection{Datasets}
\textbf{Code Search Net (CSN)\footnote{\hyperref[]{https://github.com/github/CodeSearchNet}}} \citep{Husain2019CodeSearchNetCE} is collected from publicly available open-source non-forked GitHub repositories. Only projects that are referenced by at least one other project are included.
The original paper filters around 500k code-documentation pairs for Python. They removed pairs where either the documents are less than 3 words or methods less than 3 lines. They also removed duplicate code, constructor and extension methods. After processing, there are 412k training data, 22k validation data and 22k test data. 


\textbf{Edinburgh Code-to-Docstring dataset (CDC)\footnote{\hyperref[]{https://github.com/EdinburghNLP/code-docstring-corpus}}} \citep{Barone2017APC} is a parallel Python function-to-docstring corpus collected and processed from Github. The Edinburgh Code-to-Docstring dataset contains 150,370 triples of function declarations, docstrings and bodies in the main parallel corpus. This parallel corpus is partitioned into training/ validation/ testing data, in which the training data contains 109,108 training data, 2,000 validation data and 2000 testing data.



\textbf{CodeSearchNet AdvTest (Adv)\footnote{\hyperref[]{https://github.com/microsoft/CodeXGLUE}}} \citep{Lu2021CodeXGLUEAM} is a Python dataset derived from the CodeSearchNet (CSN) corpus. The individual example in CodeSearchNet AdvTest is designed for the code search task. \citep{Lu2021CodeXGLUEAM} took the first paragraph of the docstring as the query for the corresponding Python function.  The function names and variables are replaced by special tokens, which we recover back with the original variables name.
The CodeSearchNet Advtest dataset contains 251,820 training data, 9,640 validation data, and 19,210 testing data.


\subsection{Data Processing}
In our experiment, we process our data in 3 steps. \textbf{(1)} Clean up Raw Code: All Python 2 code is converted to Python 3 using package 2to3\footnote{https://pypi.org/project/2to3/}, and all Python code styles remain consistent with package pep8\footnote{https://pypi.org/project/autopep8/}. Similar with the step of \cite{Clement2020PyMT5MT}. We also remove all invalid code samples that cannot be parsed to AST. After cleaning up the raw code, 99.92\% code data is remaining. \textbf{(2)} Remove comments and docstrings: Comments and docstrings are removed from the code, since these will not be predicted. \textbf{(3)} Replace indentation and newline: Indentation and newline is critical for generation a structured Python code. In our work, we replace them with special symbol § for Indentation and $\delta$ for newline.

\subsection{Experimental Set Up}

In both Phases, we use T5-based models. For Phase I and II, the code description is the main source inputs for the encoder. 

\textbf{Encoding Setup.} We use the AdamW optimizer for all the T5 models and assign learning rate 1e-4 for Phase I and Phase II. The training step for Phase I is kept at 75K and batch size at 32.  The training step for Phase II is kept at 100K and batch size at 32.The learning scheduler is inverse root square and has warm-up step of 5000 for phase I and 10000 for Phase II.

\textbf{Decoding Setup.} Both Phase I and Phase II take length 512 as input and have output length of 128 for phase I and 256 for Phase II. The beam size is 5 for all Phases fine-tuning.
We add repetition penalty of 2 for Syntax-Flow and Variable-Flow generation considering the case that repeated statements occur frequently.
All the tasks are run on the two Nvidia GeForce A6000 with 48GB GPU memory each.


\textbf{Evaluation.} For our experimental evaluation, we use the metrics BLEU \citep{Papineni02bleu:a}, ROUGE \citep{Lin2004ROUGEAP} and CodeBLUE \citep{Ren2020CodeBLEUAM}. BLUE and Rouge are the most common metrics to evaluate generated text. CodeBLUE is a metric specifically designed for the evaluation of generated programming languages. Apart from the similarity of the tokens, it also considers the syntax of commands and logic.

\section{Results and Analysis}
In this section, we first present our Phase I experimental results, which contain the performance of Syntax-Flow and Variable-Flow generation on the CSN, CDC, and Adv Test datasets. Then, we present our Phase II experimental results on CSN, CDC, and Adv Test datasets. We train our models on each dataset's training data, and run evaluations on the corresponding testing data. Finally, we compare our approach's performance on automatic Python code generation task with previous works.
 
\subsection{Results of Phase I}
\textbf{Syntax-Flow Results.} We first show our results on generating Syntax-Flow using T5 language model. We evaluate the generated Syntax-Flow with Rouge and BLEU metrics, as shown in Table \ref{table::phase1}. The Syntax-Flow performance of CSN, CDC, and Adv is around 50\% in Rouge-F1 and Rouge-F2, and over 35\% in Rouge-F2. These results are good considering the real vocabulary size used in Syntax-Flow is relatively smaller and syntax tokens are generally similar. When we make a comparison among the three corpora, results from CDC are slightly better than that of Adv and CSN for all the metrics consistently. CDC is a well-organized dataset that's specifically designed for Python automatic code generation task. Considering Adv is derived from CSN and thus more organized, there is only 1.3\% in Rouge score and 1\% in BLEU improvement.

\textbf{Variable-Flow Results.} We evaluate our generated Variable-Flow results from code docstrings using the Rouge and BLEU metrics. Our evaluation results regarding the generated Variable-Flow are shown in Table~\ref{table::phase1}. Similar to the results in Syntax-Flow, the performance of Variable-Flow in CDC is slightly better than the other two datasets for all the metrics scores. The average results of the Variable-Flow are not as good as those of Syntax-Flow because the generation of Variable-Flow variant components is much more difficult than the Syntax-Flow invariant components. Moreover, over 95\% of Syntax-Flow samples' lengths are shorter than 125 tokens. The Rouge F1 is over 35\% and Rouge F2 is over 15\% on average. 


\subsection{Results of Phase II}


From Table~\ref{Phase_II}, we observe the performance of final Python code generation with Rouge, BLEU and CodeBLEU. The result of CDC is the best among the three corpora because of its cleaner data as well as the effect of better Phase I performance (Syntax-Flow and Variable-Flow). CSN's results were slightly better than Adv's since CSN had about twice as much training data as Adv. As we can see from Table~\ref{other_results}, the performance of GAP-Gen slightly outperforms the T5 model that's directly trained to generate Python code for both Rouge and BLEU metrics. It indicates that our pipeline approach is effective in improving Python code generation. Similar conclusion can be proved by fine-tuning the CodeT5 language model with our GAP-Gen training pipeline. We apply our training pipeline with CodeT5 in Phase II and show that GAP-Gen CodeT5 achieves the best Rouge, BLEU and CodeBLEU scores compared with other models on the same fine-tuning task.

There is a large gap between GAP-Gen and PyMT5 on BLEU and CodeBLEU, which is because PyMT5 generates sequence with max tokens 1024. We limit the maximum target length to 256, which covers about 75\% of code lengths. Based on our comparison between T5 and GAP-Gen, the results of GAP-Gen have improved due to the pre-requisite of Syntax-Flow and Variable-Flow generation.

\subsection{Discussion}



Unlike other works focused on pre-training, we design a pipeline approach to achieve a better fine-tuning result. Given the same training configuration, our results prove that there is an improvement derived from using docstring, Syntax-Flow and Variable-Flow together, as compared to using the docstring only. Code generation is a translation task but has its own difficulties. First, our docstring inputs are usually very short, while code outputs are long. For example, there is about 85\% of the input sequences in CSN, CDC, and Adv are less than 128 tokens while over half of codes that are longer than 128 tokens. Moreover, code has stricter syntax and less ambivalent semantics. Our pipeline, by dividing the load of generating syntax and semantic information to multiple language models, bypasses the above difficulties and achieves better generation results.

The data leaking issue exists in many previous works using the pre-training technique on the automatic code generation task. For example, in previous work \citep{Clement2020PyMT5MT}, the dataset CodeSearchNet used for fine-tuning overlaps with their data used for pre-training. Both of them are collected from the public github repositories. Data leaking will tend to result in high performance on the fine-tuning task but usually is dubious in practice because model should generalize on the unseen data. In our work, we fine-tune our model using T5 which is not pre-trained on existing Code-to-Docstring datasets. Hence, T5 does not have the data leaking problem. However, CodeT5 is pre-trained on the CSN dataset, which may lead to the data leaking problem in code generation task. This can be the reason that CodeT5 alone without using our training pipeline can achieve very good results. However, after we fine-tune CodeT5 using our training pipeline, CodeT5 shows better performance on the Python code generation task, as you can see in Table~\ref{Phase_II}. 


At the same time, due to the computational resources limitation, the maximum batch size we can use is 32. Although we are limited by computational resources, we still achieve result improvements in the code generation task. In the future, better computational resources would probably increase the performance further.






\section{Conclusions}
In this work, we demonstrate the effectiveness of injecting Python syntactic and semantic information into the code generation tasks. We design and implement two different types of information components: Syntax-Flow and Variable-Flow. To incorporate this information, we encode them using separate language models and then feed them along with the docstring input into the final language model. Pre-trained language models fine-tuned with our proposed pipeline show better performances over state-of-the-art code generation models. For future directions, new strategies for incorporating that information can be explored. 


\newpage

\bibliography{anthology,custom}
\bibliographystyle{acl_natbib}

\newpage
\appendix

\section{Appendix}

\begin{table*}[ht]
\centering
\begin{tabular}{|l|c|c|c|c|c|}
\hline
                   & Rouge1-F1    & Rouge2-F1     & RougeL-F1  & BLEU   & CodeBLEU                        \\ \hline
T5                 & 30.4   & 11.7   & 27.4   & 20.7  & 21.7                                              \\ \hline
T5-Syntax-Flow   & 30.9  & 12.1  & 27.7   & 20.7  &  21.9                                               \\ \hline
T5-Variable-Flow & 30.5  & 11.8  & 27.4   & 20.6  & 21.7                                                \\ \hline
GAP-Gen-T5         & \textbf{31.1}  & \textbf{12.1}  & \textbf{27.9}   & \textbf{21.2}   & \textbf{22.1}  \\ \hline

\end{tabular}
\caption{Results Comparisons of GAP-Gen-T5 components on CSN datasets for Python code generation task.}
\label{components_results}
\end{table*}

\begin{figure*}[h!] 
\centering
\includegraphics[width=\textwidth]{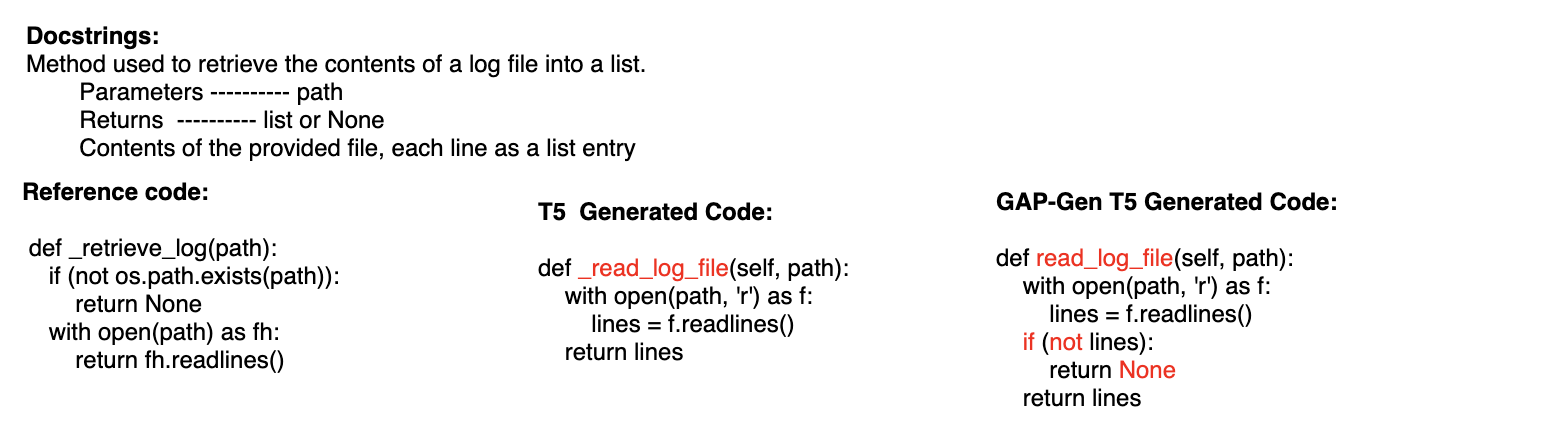}
\caption{Sample code generated from the docstring in CSN datasets. \textbf{The most left code} is the golden standard reference code. \textbf{The middle code} is generated directly from T5 fine-tuned with docstring. \textbf{The most right code} is generated using our GAP-Gen fine-tuning pipeline.}
\label{fig:example_4}
\end{figure*}

\begin{figure*}[h!] 
\centering
\includegraphics[width=\textwidth]{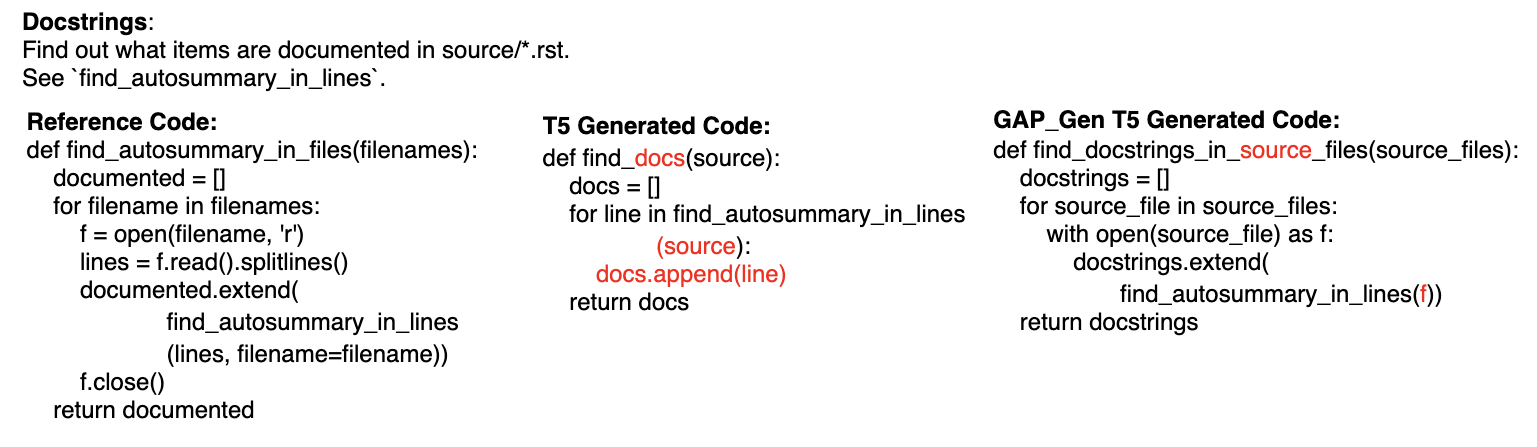}
\caption{Sample code generated from the docstring in CSN datasets.}
\label{fig:example_5}
\end{figure*}


\section{Ablation Study}


\subsection{Effectiveness of Syntax-Flow}
In order to show the effectiveness of Syntax-Flow on improving the language model's capability for the Python code generation task, we make comparisons between the results of T5 fine-tuned with docstrings only and T5 fine-tuned with docstrings and Syntax-Flow shown in Table~\ref{components_results}. Based on the comparisons of the results, we can observe that T5$_{\_{Syntax-Flow}}$ has outperformed the performance of T5 on the majority of evaluation metric scores.  Since code has a tree structure and needs to be compiled based on the corresponding AST, it is particularly important to make sure that the syntactic structure included in the generated code is correct. In T5$_{\_{Syntax-Flow}}$, we inject the syntactic structure of code, the Syntax-Flow, into the fine-tuning process of the T5 model so that the T5 can learn how the syntactic structure of code should be incorporated to generate higher quality code, a fact which we believe is the reason that T5$_{\_{Syntax-Flow}}$ has better code generation performance. 



\subsection{Effectiveness of Variable-Flow}
We also make experiments to show the effectiveness of Variable-Flow for the Python code generation task. Similarly, we make comparisons between the results of T5 fine-tuned with docstrings only and T5 fine-tuned with docstrings and Variable-Flow shown in Table~\ref{components_results}. As we can observe from the result comparison, T5$_{\_{Variable-Flow}}$ only does not achieve significant improvements regarding the evaluation metric scores and we believe that there are two potential reasons causing this to happen. First, comparing the evaluation score of the generated Syntax-Flow with that of Variable-Flow shown in Table~\ref{table::phase1}, we can see that the generated Variable-Flow's evaluation scores are worse than that of the Syntax-Flow. It happens because the length of generated Variable-Flow is much longer than that of Syntax-Flow due to the characteristics of Variable-Flow that it supposes to contain general semantic information of code. Second, due to the longer length of the generated Variable-Flow, the inputs to T5$_{\_{Variable-Flow}}$ are much longer than that of T5$_{\_{Syntax-Flow}}$, and T5$_{\_{Variable-Flow}}$ does not know which line of generated code the variable should be assigned to because of the lack of syntactic structure information, then much longer code is likely to be generated. However, the effectiveness of Variable-Flow can also be reflected from the loss of T5 with Variable-Flow only from Figure~\ref{fig:t5-gap-gen-var}. A lower perplexity can be obtained, and the generated code is more fluent on average.

\subsection{Samples Analysis}
To illustrate the usefulness of our proposed Syntax-Flow and Variable-Flow components, we have attached the generated Python code samples using Syntax-Flow and Variable-Flow with the corresponding docstrings in the CSN dataset. We have also provided sample analysis of them in the following paragraphs. 

As we have demonstrated in our paper, in order to generate well-working Python code, the language model should not only understand the text semantic information from a given docstring but also should be capable of considering the code syntactic information and the code variable semantic information. 
 
Based on the given sample codes shown in Figure~\ref{fig:example_4}, it is clear that the code, which is generated directly from T5 without having Syntax-Flow and Variable-Flow injected, cannot properly handle both the code syntactic information and the code variable semantic information. For example, the docstring specifies that the code should return a list or None variable, suggesting that there are 2 different return values that should be generated under different conditions. As a result, the fine-tuned model should consider both the code syntax logic, the boolean operation, and the variable semantic, the generated variables, during the code generation process. However, due to the lack of Syntax-Flow and Variable-Flow components, the T5 model fine-tuned with docstring only is unable to learn the code syntactic information and the code variable semantic information, resulting in the fine-tuned model generates code that is not able to determine where the boolean operation should be generated to handle multiple return values. Similar trends happen in the sample codes shown in Figure~\ref{fig:example_5} as well.

In contrast, in our work GAP-Gen, we consider the Syntax-Flow and Variable-Flow during the code generation process. Due to the support of these two components, we can successfully generate a higher-quality code with the boolean operation and different return values.


\subsection{Training Algorithms}
In this subsection, we include the training algorithms for \textbf{1.} generating the Syntax-Flow and Variable-Flow, and \textbf{2.} fine-tuning the pre-trained Language Model with Syntax-Flow and Variable-Flow.
\label{sec:appendix}
\begin{algorithm}[ht!]
\caption{Generate Syntax-Flow \& Variable-Flow }\label{alg:wordy}
\begin{algorithmic}[1]
\Require{
\Statex $x = (x_1,x_2,...,x_n) \in X$: input docstring
\Statex $LM_S$: Language Model being used for Syntax-Flow
\Statex $LM_V$: Language Model being used for Variable-Flow 
}
\Ensure{\Statex $A = (a_1,a_2,...,a_n)$: Syntax-Flow
\Statex $B = (b_1,b_2,...,b_k)$: k Variable-Flow }
 \State Initialize A, B to be empty arrays
 \For{each docstring:$x \in X$}
  \State $a \gets LM_S(x)$
  \State $b \gets LM_V(x)$
  \State Append(A,a)
  \State Append(A,a)
 \EndFor
\State \Return A,B
\end{algorithmic}
\end{algorithm}

\begin{algorithm}[ht!]
\caption{Fine-tuning Language Model with Syntax-Flow}\label{alg:wordy}
\begin{algorithmic}[1]
\Require{
\Statex $x = (x_1,x_2,...,x_n) \in X$: input docstring
\Statex $LM_S$: Pre-trained Language Model being used for Syntax-Flow
}
\Ensure{\Statex $LM_S$: Language Model fine-tuned for generating Syntax-Flow }
 \State Initialize D to be empty array
 \For{each docstring:$x \in X$}
  \State $p \gets \textproc{ASTParse(x)}$ AST parsed by standard Python AST parser
  \State $d \gets \textproc{SynParse}(p)$
  \State Append(D,d)
 \EndFor
 \For{i = 1 to |X|}
  \State $a' \gets LM_S(X[i])$ 
  \State $l = loss(D[i],a')$
  \State $LM_S.backwards(l)$
 \EndFor
\State \Return $LM_S$
\end{algorithmic}
\end{algorithm}

\subsection{Code Generation Loss Analysis}
\begin{figure*}[ht] 
\centering
\includegraphics[width=\textwidth]{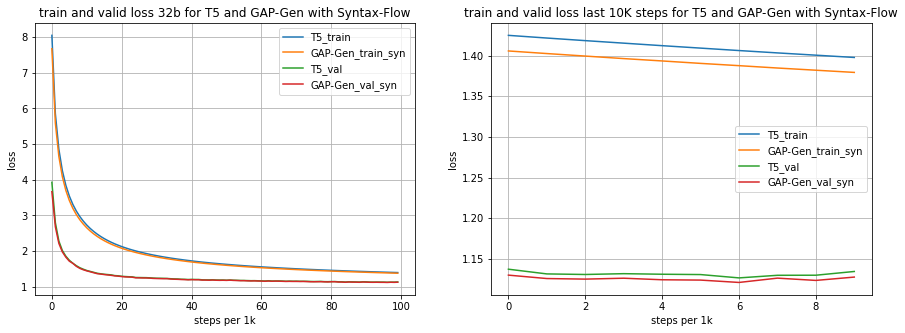}
\vspace{-10pt}
\caption{Loss comparison visualization between T5 and GAP-Gen using Syntax-Flow only.}
\label{fig:t5-gap-gen-syn}
\end{figure*}

\begin{figure*}[ht] 
\centering
\includegraphics[width=\textwidth]{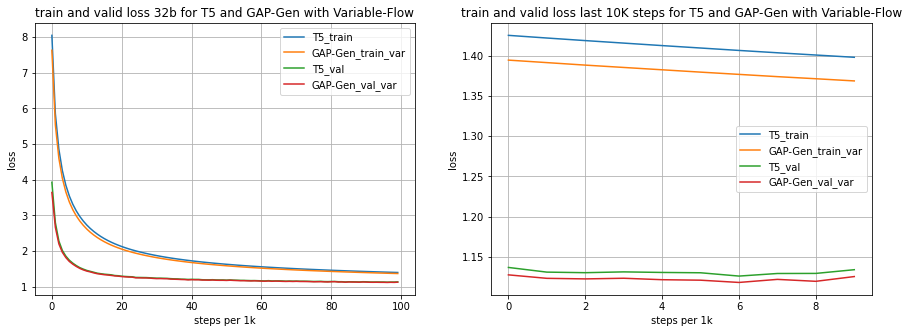}
\vspace{-10pt}
\caption{Loss comparison visualization between T5 and GAP-Gen using Variable-Flow only.}
\label{fig:t5-gap-gen-var}
\end{figure*}

We further analyze the loss trend for generating the Python code using T5 and our GAP-Gen training pipeline. In our analysis, we show three loss trend comparisons: 
\begin{itemize}
    \item T5 vs GAP-Gen with Syntax-Flow only in Figure~\ref{fig:t5-gap-gen-syn},
    \item T5 vs GAP-Gen with Variable-Flow only in Figure~\ref{fig:t5-gap-gen-var},
    \item T5 vs GAP-Gen with both Syntax-Flow and Variable-Flow in Figure~\ref{fig:t5-gap-gen}.
\end{itemize}

Based on our observations, we find the global loss trends between T5 and GAP-Gen are similar. However, when we zoom into the last 10k steps, the training and validation loss of GAP-Gen are consistently lower than those of T5 on the Python code generation task from all three scenarios. 

By comparing with the loss trend between T5 and GAP-Gen with Syntax-Flow only and GAP-Gen with Variable-Flow only, we find both scenarios have lower training and validation loss in the last 10k steps than those of T5. This fact shows that both of Syntax-Flow and Variable-Flow are contributing to the Language Model fine-tuning process. At the same time, we find GAP-Gen with both the Variable-Flow and Syntax-Flow results in the lowest training and validation loss compared with those of the other two scenarios. This finding further illustrates that our method GAP-Gen does have improvement on the Python code generation task. 

\begin{figure*}[ht] 
\centering
\includegraphics[width=\textwidth]{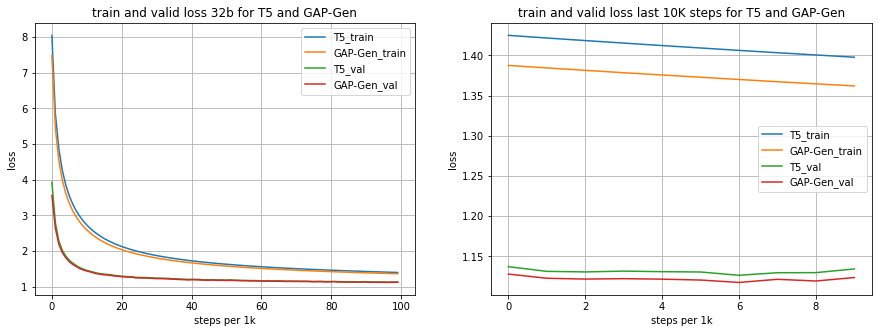}
\vspace{-10pt}
\caption{Loss comparison visualization between T5 and GAP-Gen.}
\label{fig:t5-gap-gen}
\end{figure*}

\begin{figure*}[ht] 
\includegraphics[width=\linewidth]{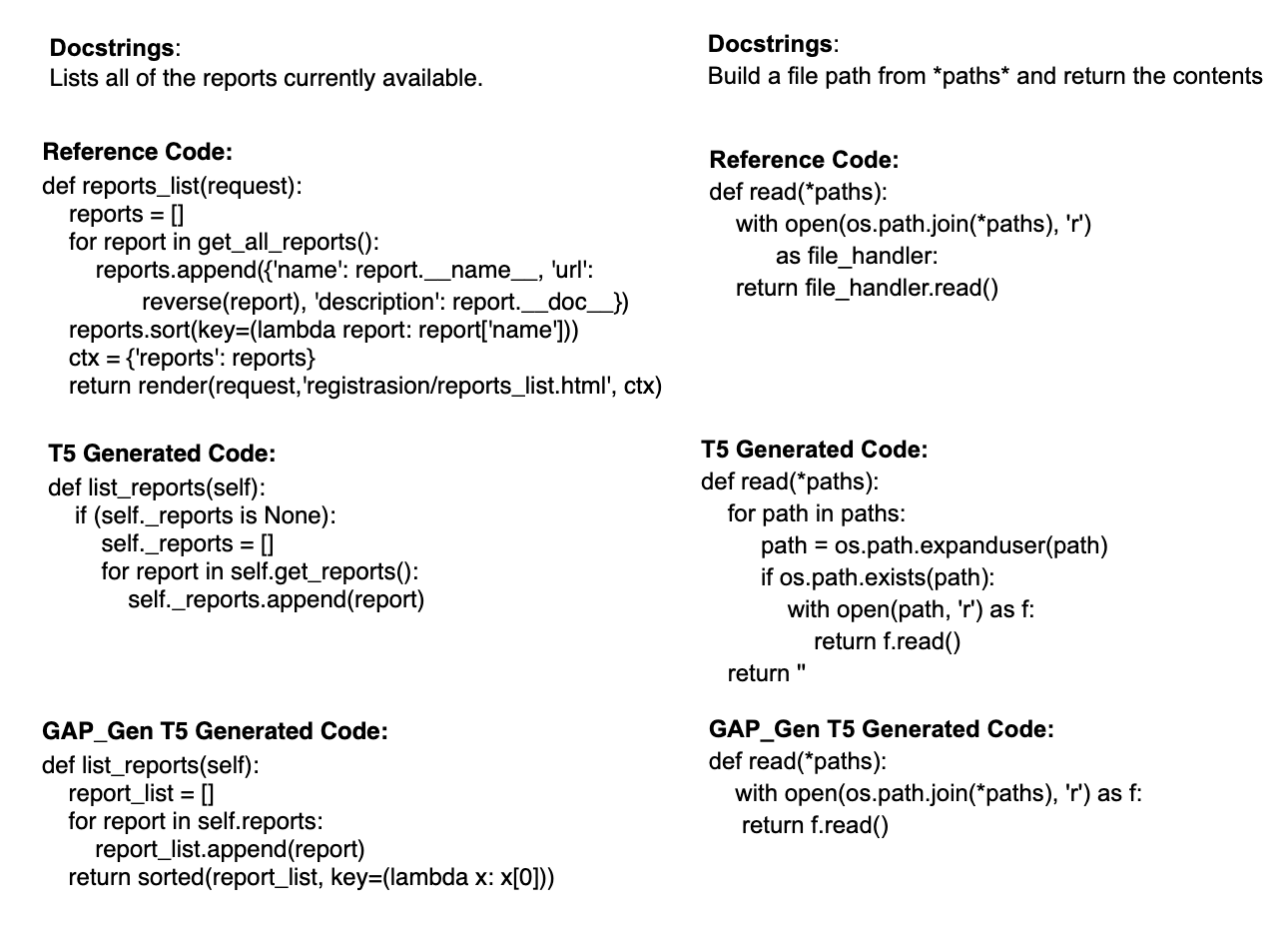}
\includegraphics[width=0.8\linewidth]{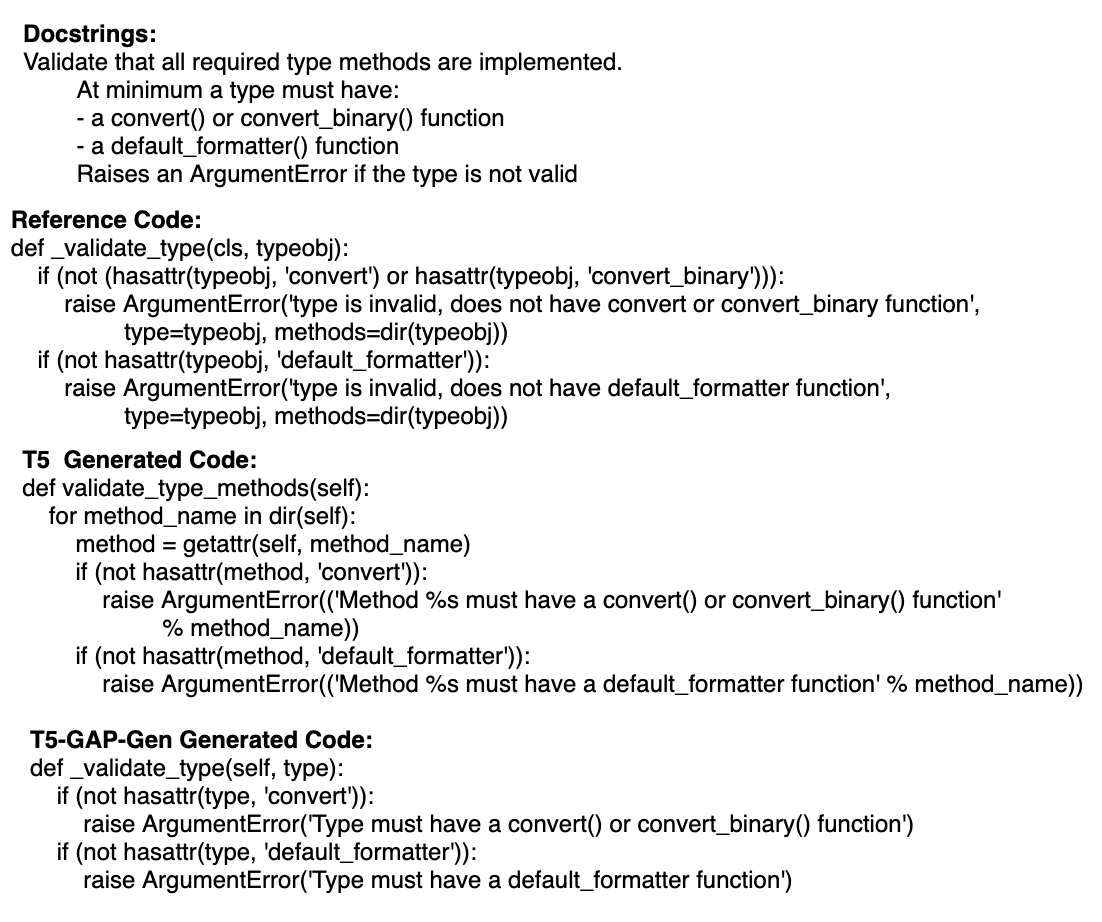}
\vspace{-10pt}
\caption{Additional generated sample codes from our experiments.}
\label{fig:t5-gap-gen}
\end{figure*}

\end{document}